\documentclass[11pt,twoside]{article}
\usepackage{asp2010}
\newcommand{\lsim}{\, \lower2truept\hbox{${< \atop\hbox{\raise4truept\hbox{$\sim$}}}$}\,}
\newcommand{\gsim}{\, \lower2truept\hbox{${> \atop\hbox{\raise4truept\hbox{$\sim$}}}$}\,}
\newcommand{\solar}{\odot}

\resetcounters

\begin{document}

\title{NIR/Optical Selected Local Mergers --- 
Spatial Density and sSFR Enhancement}
\author{C. Kevin Xu$^1$}
\affil{$^1$ Infrared Processing and Analysis Center, 
California Institute of Technology 100-22, Pasadena, CA 91125, USA}

\begin{abstract}
Mergers play important roles in triggering the most active
objects in the universe, including (U)LIRGs and QSOs. However, whether
they are also important for the total stellar mass build-up in
galaxies in general is unclear and controversial. The answer to that
question depends on the merger rate and the average strength of
merger induced star formation.  In this talk, I will review studies on
spatial density and sSFR enhancement of local mergers found in
NIR/optical selected pair samples. In line with the current literature
on galaxy formation/evolution, special attention will be paid to the
dependence of the local merger rate and of the sSFR enhancement on four
fundamental observables: (1) stellar mass, (2) mass ratio, (3) separation, 
and (4) environment.
\end{abstract}

\section{Introduction}
Historically, galaxy mergers were associated with ``pathological
galaxies'' \citep{Shapley1943} that did not fit into the ``Hubble
Sequence''.  Simulations of \citet{Toomre1972} established firmly
the connection between gravitational tidal effects during galaxy
merging and morphological features such as long tails, bridges and
plumes, seen abundantly in ``Atlas of Peculiar Galaxies''
\citep{Arp1966}.  Later studies found evidence for mergers to play
important roles in triggering extreme starbursts (such as ULIRGs,
\citealt{Sanders1996}) and QSOs 
\citep{Heckman1986, Sanders1988, Bahcall1995}.  
Alar \citet{Toomre1978} proposed that merging can
transform spirals into ellipticals, a theory that puts mergers on the
center stage of galaxy evolution. In the hierarchical structure
formation paradigm of the contemporary cosmology, galaxy and dark
matter halo (DMH) merging is one of the most significant processes
affecting the evolution of structures in the early universe, and is
largely responsible for the growth of massive dark matter halos and
the buildup of galaxies \citep{Kauffmann1993, Lacey1993,
  Khochfar2005}.  In cosmological simulations, it is often
assumed that merger induced star formation (SF) is the major (or even
the dominant) contributor to the high star formation rate (SFR) at $z
\sim 1$ -- 2, the epoch when the net SFR in the universe peaks
\citep{Guiderdoni1998, Somerville2001, Baugh2005}.  On
the other hand, observationally, we are still facing two unsettled
questions: (1) how many galaxies
are involved in mergers? (2) is there any significant SFR enhancement
in an average merging galaxy (which is very different from FIR selected
mergers such as ULIRGs)?
In this talk I will review the literature on studies of these 
two topics for local mergers ($z \lsim 0.1$). Studies of mergers
of higher redshifts will be covered elsewhere in this conference.

\section{Merger Fraction in the Local Universe}

Mergers are selected either by counting galaxies in close pairs that
are destined to merge, or by identifying peculiar galaxies with
morphological features associated to mergers. Generally,
the formal selects early stage mergers when the two galaxies are
still visually separable,
and the latter favors later stage mergers (or even post-mergers)
because the tidal features is not fully developed until
several $10^8$ years after the first close-encounter, and lasts
well after the coalescence \citep{Barnes1996}. A summary of
results on the local merger fraction $\rm f_{mg}$ is listed in Table~1.

\noindent
\begin{table}[!ht]
\vspace*{-8mm}
\caption{Results on Local Merger Fraction $\rm f_{mg}$}
\vspace*{-2mm}
\begin{center}
{
\footnotesize
\begin{tabular}{llcccccc}
\tableline
\noalign{\smallskip}
 reference &   method    &   selection
&   $\rm \mu_{max}$  &   $r_{max}$
&   $\Delta v_{max}$ &  $\rm N_{mg}$ 
&   $\rm f_{mg}$ \\ 
&                     &    
&                      &   $\rm h^{-1}\, kpc$ 
&   $\rm km\; sec^{-1}$  &    
&  \%     \\           
\noalign{\smallskip}
\tableline
\noalign{\smallskip}
\citet{Fried1988} & morph & $\rm B < 12.0$ & ...  & ... & ... & 3 &  3 \\
\citet{Xu1991} & pair & $\rm m_{Zw} \leq 15.6$  & ... & ... & 1000 & 921 & 10 \\
\citet{Patton1997} & pair & $\rm m_{ph} \leq 14.5$ & ... & 20 & 500 & 130 & $\rm 4.3 \pm 0.4 $ \\
\citet{Patton2000} & pair &  $\rm -21 < M_{B} < -18$ & ...
& 20 & 500 & 80 & $\rm 2.3 \pm 0.5$ \\
\citet{Xu2004} & pair & $\rm K\leq 12.5$ & 2.5 & 20 & 500 & 30 & $\rm 1.7\pm 0.3$ \\
\citet{Bell2006} & 2pcf & $\rm M_{star}> 2.5\times 10^{10} M_\solar$ 
& ... & 21 &... &... & $\rm 1.1\pm 0.2$ \\
\citet{dePropris2007} & pair & $\rm -21 < M_{B} -5\log{h} < -18$ & ... & 20 & 500 
& 112 & $\rm 4.1\pm 0.4$ \\
\citet{Kartaltepe2007} & pair & $\rm M_{V} < -19.7$ & ...
& 20 & 500 & 90 & $\rm 0.7\pm 0.1$ \\
\citet{Patton2008} & pair & $\rm 14.5\leq m_{r} \leq 17.5$ & 2.0 & 20 & 500& 473 & $\rm 2.1\pm 0.1$ \\
\citet{Domingue2009} & pair & $\rm K\leq 12.5$ & 2.5 & 20 & 1000 & 265 & $\rm 1.6\pm 0.1$ \\
\citet{Robaina2010} & 2pcf & $\rm M_{star}> 5\times 10^{10} M_\solar$ 
& ... & 21 &... &... & $\rm 1.4\pm 0.2$ \\
\citet{Darg2010} & morph & $\rm M_{r} < -20.55$ & 3.0 & ... & 500 & 1243 &  $\rm 3.0\pm 1.5$ \\
\citet{Xu2012} & pair & $\rm K\leq 12.5$ & 2.5 & 20 & 500 & 221 & $\rm 1.3\pm 0.1$ \\
\noalign{\smallskip}
\tableline
\end{tabular}
}
\end{center}
\end{table}

\vspace*{-5mm}
The early result of \citet{Xu1991} is high ($\rm f_{mg} = 10\%$)
because it includes wide pairs (up to projected separation of $\rm
r \sim 100\; kpc$) and both major and minor mergers. Pair samples of
\citet{Patton1997}, \citet{Patton2000}, 
\citet{dePropris2007} and \citet{Kartaltepe2007} are
absolute magnitude limited without any restriction on the mass ratio
or the luminosity ratio. It is difficult to compare these results with
any model because the mass ratio is one of the most important
parameters in merger models. Also, these pair samples suffer from the
``missing secondary'' incompleteness, namely many paired galaxies are
missed because their companions are fainter than the absolute
magnitude limit.  This bias is avoided in the works of \citet{Xu2004},
  \citet{Patton2008}, \citet{Domingue2009} and 
\citet{Xu2012}, who confined their samples to
close major-merger pairs of 
projected separation $\rm r \leq 20\; h^{-1}kpc$
and $\rm \mu \leq 2.5$ or $\rm \mu \leq 2$, 
$\mu$ being the primary-to-secondary mass ratio
($\rm \mu = M_{pri}/M_{2nd}$). \citet{Xu2012} is
an update of \citet{Xu2004} and \citet{Domingue2009},
with two major improvements: (1) a correction for the contamination of
unphysical pairs caused by galaxy clustering; (2) exclusion of the
local super-cluster region.
The result of \citet{Patton2008} is higher than that of
\citet{Xu2012}, and the difference is more significant after adjusting
the different values of $\rm \mu_{max}$.  Both studies used galaxies
taken from the SDSS database.
\citet{Patton2008} included only pairs with both components having
SDSS redshifts.  This resulted in a rather low redshift completeness of
$\sim 30\%$ for the pairs because of the
``fiber collision'' issue in the SDSS redshift survey. The large
correction factor ($\gsim 3$) for this incompleteness may indeed
introduce large uncertainties in the result.  In
contrast, \citet{Xu2012} included pairs with at least one SDSS
redshift, and made their own redshift observations and literature
searches for the missing redshifts. The completeness of this sample,
only missing pairs with neither component having SDSS redshift, is
$89\%$ \citep{Domingue2009}. Another major difference between
\citet{Patton2008} and \citet{Xu2012} is that the former was selected
in the optical r-band while the latter in the NIR K-band. The merger
induced SF may boost the $\rm L_{r}$ 
(which includes the $\rm H_\alpha$ emission) of paired galaxies
and in turn boost the $f_{mg}$ for a given $\rm L_{r}$ because of the
steep decline of the luminosity function. On the other hand, dominated
by the emission of old stars,  $\rm L_{K}$ is insensitive to
the enhanced SFR.

The seemingly good agreement between the early result of
\citet{Fried1988} and the recent result of \citet{Darg2010} for
morphologically selected mergers is fortuitous. The sample of
\citet{Fried1988} includes both major mergers and minor mergers, while
that of \citet{Darg2010} includes only the major mergers ($\rm \mu
\leq 3$). Excluding peculiar galaxies that cannot
be resolved into two galaxies, the sample of \citet{Darg2010}
is incomplete for
morphologically selected mergers. Nevertheless, the agreement between
merger fractions derived using morphologically selected merger samples
and using close pair samples ($\rm r \leq 20\; h^{-1}kpc$) is
remarkable, indicating that the merger time scales for these two different
selections may indeed be similar \citep{Lotz2010, Conselice2009, Xu2012}.

In addition to morphological and pair selected mergers, I also
included in Table~1 studies based on 2-point correlation functions
(2pcf). \citet{Bell2006} and \citet{Robaina2010} have shown that
for massive galaxies the 2pcf
can be well extrapolated down to $\rm r = 15\; kpc$, as power-laws
with $\rm \gamma \simeq 2$. Thus, the pair fraction can be estimated
from the 2pcf as following:
\begin{equation}
\rm f_{mg} = 4\pi n \int^{r_{max}}_{r_{min}} \, [1+\xi (r)] r^2 dr, \label{eq:2pcf}
\end{equation}
where $n$ is the number density of galaxies and $\rm \xi$ the 2pcf.
The results are indeed in good agreement with those from pair counts.
On the other hand, similar to those derived using absolute magnitude limited
pair samples, these results suffer from a lack of the mass ratio information.

For close major-merger pairs, \citet{Patton2008}, \citet{Domingue2009} and
\citet{Xu2012} found that $\rm f_{mg}$ is constant against the
luminosity or the stellar mass $\rm M_{star}$. \citet{Xu2004} found a
positive dependence of $\rm f_{mg}$ on $\rm M_{star}$, but that result
has a large uncertainty due to the small sample size (19 pairs). Much
of the discrepancies between the results derived using pairs, as
listed in Table~1, can be attributed to the differences in the
separation and the mass ratio.  The separation dependence of $\rm
f_{mg}$ can be derived using Eq.~\ref{eq:2pcf}, assuming $\rm \xi (r)
= (r_0/r)^\gamma$ and $\rm \gamma=2$. This results in a $\rm
f_{mg}\propto r_{max}$ (assuming $\rm r_{min} << r_{max}$). 
\citet{Xu2012} found that, within the range of $\rm
1 \leq \mu_{max} \leq 10$, $\rm f_{mg}$ is approximately proportional to
$\rm \log (\mu_{max})$. This suggests that there are about
equal numbers of major mergers with $\rm 1 \leq \mu \leq 3$ and
minor mergers with $\rm 3 \leq \mu \leq 10$, contradicting
a common belief that such minor mergers are much more 
abundant than major mergers! Extrapolating the result
of \citet{Xu2012} using these $\rm r$- and $\rm \mu_{max}$- dependences, we
have:
\begin{equation}
\rm f_{mg} = 1.5\% \times {r_{max}\over 20\; h^{-1}kpc} 
\times {\log(\mu_{max}) \over \log(3.0)}.
\end{equation}
According to \citet{Ellison2010}, both the average
separation and the average velocity difference of pairs increase with
the local density $n$, while the average mass ratio is insensitive to $n$.
\citet{Xu2012} found that $90.4 \pm 2.5\%$ of
K-band selected close major-merger pairs in the sample of \citet{Domingue2009},
with $\Delta v_{max} = 1000\; km\; sec^{-1}$,
have $\Delta v \leq 500\; km\; sec^{-1}$.  

The differential merger rate $\rm R_{mg}$ is the probability for
each galaxy to be involved in a major merger per Gyr:
$\rm R_{mg} = f_{mg}/T_{mg}$, where $\rm T_{mg}$ is the time scale
(in Gyr) for the merger selection (e.g. for morphologically selected
mergers, $\rm T_{mg}$ is the time during which the tidal features are
recognizable).  It is worth noting that (1) because galaxy merger is a
very complex process \citep{Hopkins2010b}, $\rm R_{mg}$, $\rm f_{mg}$
and $\rm T_{mg}$ are all functions of the redshift, stellar mass, 
mass ratio, separation,
and other parameters; (2) for a given $\rm R_{mg}$, the merger fraction
$\rm f_{mg}$ can be different in different merger selections because
of different values of $\rm T_{mg}$; (3) $\rm T_{mg}$ is one of
the major sources of uncertainties in the calculation of merger rate;
for example, for close major-merger pairs of $\rm L^*$ galaxies with
$\rm \mu \leq 3$ and projected separation $\rm r \leq 20\; h^{-1}kpc$,
$\rm T_{mg}$ ranges from $\sim 0.3$ Gyr \citep{Lotz2010} to
$\sim 0.9$ Gyr \citep{Kitzbichler2008}; (4) it is important
to distinguish $\rm T_{mg}$ from the total merging time scale that
starts when a companion falls into the dark matter halo (DMH) of the
target galaxy and ends when two galaxies coalesce; (5) 
many wide pairs ($\rm r \gsim 50\; h^{-1}kpc$), in particular minor mergers,
may never merge \citep{Patton2000, Lotz2010}.

\section{The sSFR Enhancement in Local Mergers}

It has been well established and well documented that, in the local universe,
the extreme starbursts such as ULIRGs 
are triggered by galaxy mergers \citep{Sanders1996}. However, there has
been a long debate on whether every merging galaxy
(or, in a weaker version, most merging galaxies)
has significantly enhanced SF activity, presumably
triggered by the gravitation tidal effect and other effects (e.g. 
enhanced collision rate of gas clouds) associated with merger.

Merger induced star formation was first predicted by
\citet{Toomre1972}, and confirmed by  \citet{Larson1978} 
in a study of optical colors of Arp galaxies. Many
subsequent studies of the $\rm H_{\alpha}$ emission and FIR
emission (both are SFR indicators) 
in Arp galaxies and in paired galaxies provided further support
to this theory (see \citealt{Kennicutt1996} for a review). On the other
hand, \citet{Haynes1988} found little or no enhanced FIR emission 
in a
sample of optically selected pairs compared to a control sample of
single galaxies.  In a more influential paper, 
\citet{Bergvall2003} reported a multi-wavelength study in which
they found no significant
SFR enhancement for a sample of morphologically selected merger
candidates. Apparently, only some merging
galaxies have significantly enhanced SFR (with ULIRGs as the extreme
examples) and the others do not. 
{\bf Whether the mean SFR of a merger sample shows
significant enhancement depends very much on how the sample is
selected.} \citet{Kennicutt1987} and \citet{Bushouse1988} found that
merger candidates which show strong signs of tidal interactions have
significantly stronger SFR enhancement than optically selected paired
galaxies, the latter being only marginally enhanced (a factor of
$\sim 2$) compared to single galaxies.  \citet{Telesco1988} found a
strong tendency for pairs with the highest far-IR color temperatures
to have the smallest separation. \citet{Xu1991} showed that
the enhancement of the FIR emission 
of close spiral-spiral (S+S) pairs with separation less
than the size of the primary and with signs of interaction is significantly
stronger than that of wider pairs and pairs without interaction signs.
\citet{Sulentic1989} found that elliptical-elliptical (E+E) pairs are
equally quiet in the FIR emission as single ellipticals.
Very few E's in S+E pairs are FIR bright,
possibly cross-fueled by their S companions \citep{Domingue2003}.

More recently, large digitized surveys (e.g. SDSS, 2MASS, 2df, etc)
enabled large and homogeneously selected pair samples. A clear
anti-correlation between the specific SFR (sSFR=SFR/$\rm M_{star}$)
and the pair separation has been well established \citep{Barton2000,
  Lambas2003, Alonso2004, Nikolic2004, Li2008, Ellison2008}. These
results also showed evidence for a threshold separation at $\rm r_{clo} =
20\, h^{-1} kpc$ (or $\rm r_{clo} = 30\, h_{70}^{-1} kpc$), beyond which
significant sSFR enhancement (i.e. no less than $\sim 30\%$,
\citealt{Ellison2008}) is not detected.

Does every star forming galaxy (SFG)
in close pairs with $\rm r \leq r_{clo}$ have enhanced sSFR? This
question was addressed by \citet{Xu2010} for major-merger pairs of
$\rm \mu \leq 2.5$.  They observed with Spitzer a complete sample of
27 K-band selected close major-merger pairs, including 39
spirals (classified according to the morphology) that do not harbor known
AGNs. Two of their results stand out: 
(1) on average, spirals in S+E pairs do
not show any sSFR enhancement compared to their
counterparts in a mass-matched control sample of single spirals; 
(2) the sSFR enhancement of spirals in S+S pairs is mass dependent in the
sense that significant sSFR enhancement (a factor of $\sim 3$) 
is confined to massive spirals
($\rm M_{star} \gsim 10^{10.4}\; M_\solar$ for a Salpeter IMF, or $\rm
M_{star} \gsim 10^{10.0}\; M_\solar$ for a Chabrier IMF) while no sSFR
enhancement is found for less massive paired spirals.
The result (1) is somewhat surprising because, if the merger induced 
SF is purely a gravitational phenomenon, one does not
expect the morphology of the companion should make any difference.
On the other hand, as pointed out by \citet{Struck2005}, the
merger induced SFR depends on two things: (i) the amount of cold gas available,
(ii) the amplitude of the gas compression (as depicted by
the ``Kennicutt-Schmidt Law'').
Because a spiral companion and an elliptical companion of the
same mass should trigger same gravitational tidal squeeze, the difference
between S in S+E pairs and S in S+S pairs must be due to the difference
in their cold gas abundance. Unfortunately this cannot be verified because
no data are available for the gas mass in these galaxies. There are
several plausible explanations for the result (2). First of all it is
likely that the threshold separation for the sSFR enhancement, $\rm
r_{clo}$, is related to the tidal radius which in turn scales with
$\rm M^{1/3}$ for major mergers. Therefore many low mass galaxies in
the sample may not be ``close'' mergers although they have $\rm r \leq
20 h^{-1} kpc$. Secondly, according to the theory proposed by \citet{Mihos1997},
low mass galaxies do not have sufficient disk
self-gravity to amplify dynamical instabilities, and this disk
stability in turn inhibits interaction-driven gas inflow and starburst
activity.  Thirdly, low mass spirals have systematically higher
gas fraction ($\rm f_{gas} \gsim 0.4$). \citet{Hopkins2009} pointed
out that in a high $\rm f_{gas}$ galaxy the merger induced
gravitational torque is inefficient in removing the angular momentum
from the cold gas and therefore is unable to transport large amount of
gas from disk to nucleus.  As a consequence, nuclear starbursts (a
major mode of merger induced SF) may be largely missing in the low
mass spirals involved in major mergers. It is worth noting that
merger samples selected in either the blue band or the $\rm H_{\alpha}$ emission
are biased for low mass late-type galaxies, and many of them indeed do
not show significant sSFR enhancement 
(e.g. \citealt{Bergvall2003, Knapen2009}).

\begin{figure}[!htb]
\plottwo{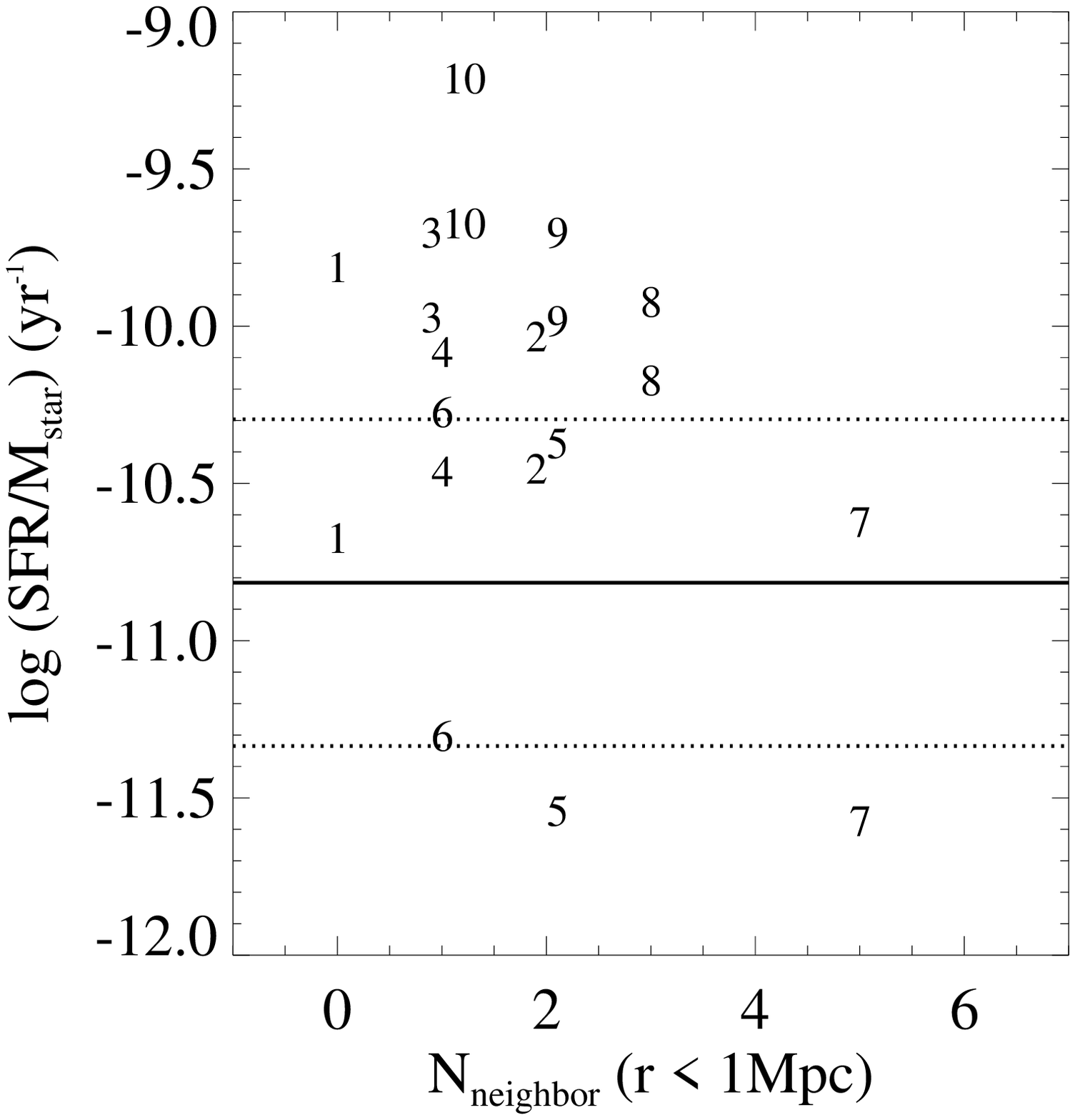}{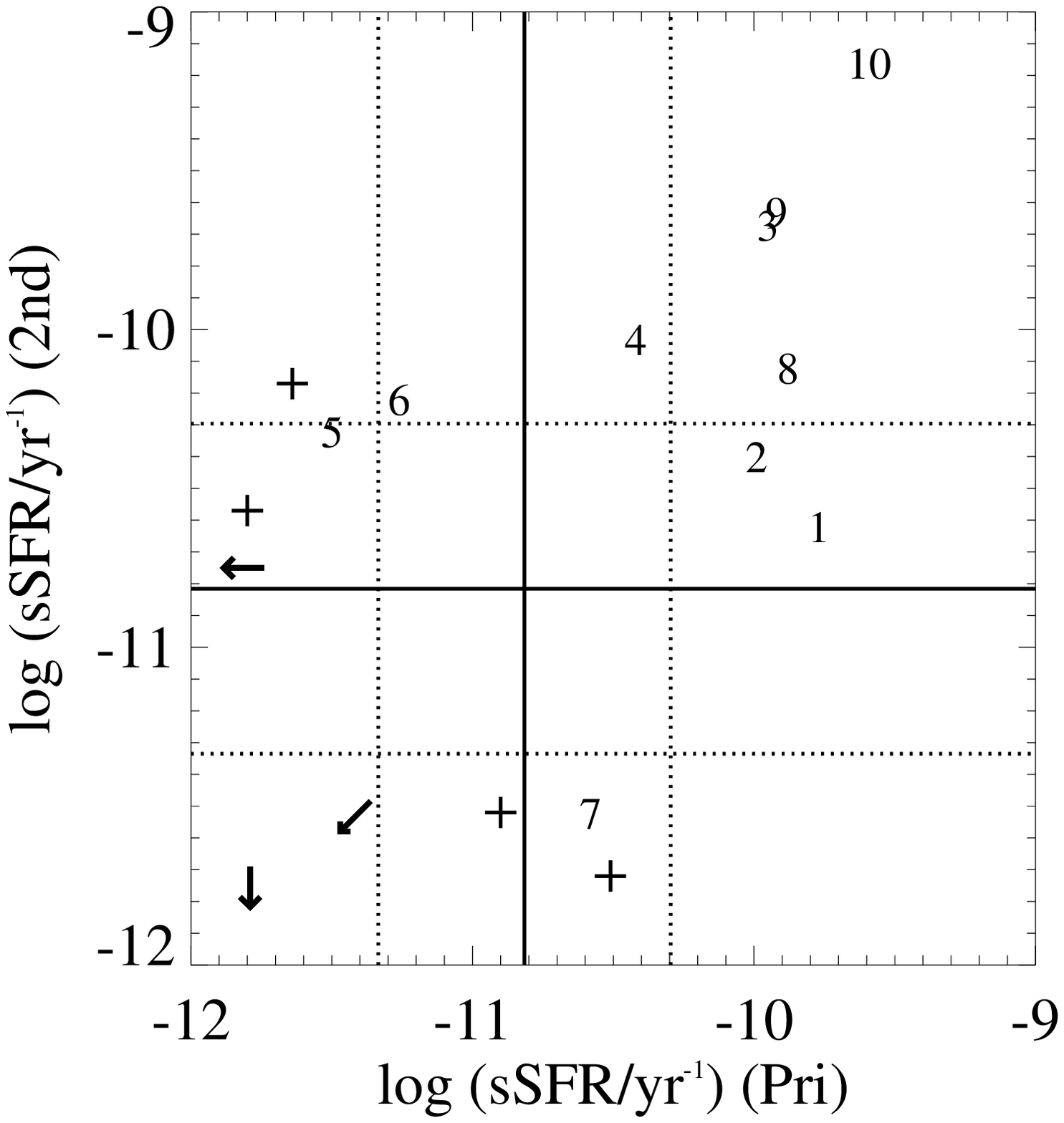}
\caption{{\bf Left (panel a)}: sSFR vs. $\rm N_{neighbor}$ plot for massive 
galaxies ($\rm M_{star} = 10^{10.7\hbox{--}11.2}\; M_\solar$, or
$\rm M_{star} = 10^{10.3\hbox{--}10.8}\; M_\solar$ for a Chabrier IMF)
in 10 S+S pairs. The two components in each pair are marked by the same
number. The solid line marks the average
sSFR of single spirals in a control sample. 
The two dotted lines mark the $\rm \pm 1\, \sigma$ 
boundaries. {\bf Right (panel b)}: Plot for the correlation between
the sSFR of two components in pairs. The numbers represent the same
S+S pairs in panel a. The crosses and upper/lower limits are for
S+E pairs. The solid lines and dotted lines have the same meaning as in
panel a.
}
\end{figure}
Another interesting result of \citet{Xu2010} is a significant
correlation (at 92\% confidence level) between the sSFR of the two
components in massive S+S pairs. This is related to the above result
(1) because, apparently, a paired galaxy knows not only about the
companion's morphology, but also its sSFR!  The correlation is not due
to the mass dependence of the sSFR because 
the pairs studied have a narrow
mass range of $\rm \Delta \log(M_{star})) \leq 0.5$,
comparable to that of the mass ratio of pairs 
($\rm \log(\mu) \leq 0.4$).  \citet{Xu2010}
did not find any significant
dependence of the sSFR on the density of local neighbors within 2 Mpc
radius, either. However,  after reducing the radius of the neighbor
searching to 1 Mpc, I did find a much
more significant dependence (Fig.~1a). This is in agreement with
\citet{Ellison2010} who found evidence for non-enhancement 
of sSFR among close pairs in high local density regions.
Nevertheless, the correlation between the sSFR of the two galaxies
in th same pair cannot be explained by this local density dependence
because the difference between the sSFR of the two
components in a given pair is much smaller than the dispersion of the
sSFR of galaxies with a given $\rm N_{neighbor}$ (Fig.~1a).  As shown in
Fig.~1b, the correlation is mainly due to pairs being separated into
two groups: a group of 7 S+S pairs with both components having relatively
high sSFR, and another group of 3 S+S pairs plus 7 S+E pairs where at
least one of the two components is a ``red-and-dead'' galaxy (either an
elliptical in case of a S+E pair or a red S0/Sa in case of a S+S pair). 
\citet{Kennicutt1987} found a similar correlation between the $\rm
H_\alpha$ emission of two components in galaxy pairs
(the so called ``Holmberg effect''). They argued that
the correlation is due to the common dependence of the SFR of both
components on some orbital parameters (e.g. $\Delta v$ and separation)
and on merger stages. However, because having a red component is
related neither to a pair's orbit nor to its merger stage, Kennicutt's
interpretation cannot explain our result. Alternatively, we propose that
the correlation is due to the modulation of the sSFR by the IGM in the
dark matter halo (DMH) that the two galaxies share (being
in the final stage of a merger, they have been within each other's
virial radius for $\rm \gsim 1 Gyr$). For example,
when the DMH has strong ``cold flows'' \citep{Dekel2009},
both galaxies may have abundant cold gas supply and therefore 
higher sSFR. On the other hand, DMH's containing
red galaxies may have little ``cold flows'', and therefore
starve the merger induced SF in 
the SFG componions. Recently, \citet{Hwang2011}
reported a similar difference between the sSFR of late-type galaxies 
with late-type neighbors and that of those with early-type neighbors 
in an FIR selected sample. They interpreted
the non-enhancement in latter case as a consequence of the SF quenching
by the hot-gas halo associated with the early-type companion. 
It is not clear whether this theory applies to our result, given that
some of our pairs in the lower-left corner of Fig.~1b do not contain
ellipticals but red spirals, which very rarely have extended
hot gas halos. Interestingly,  as shown in Fig.~1a, 
low sSFR S+S pairs \#5 and \#6 are in fairly
low local ($r< 1$ Mpc) density regions. Therefore the 
low sSFR in these two pairs might indeed be unrelated to any
effect associated with the hot gas
because lower $\rm N_{neighbor}$ usually indicates lower DMH
mass and therefore lower IGM gas temperature and lower hot gas abundance.

Late-type pairs with $\rm \Delta v \gsim 400\; km\; sec^{-1}$, most
residing in dense groups or clusters \citep{Domingue2009}, 
do not show any sSFR enhancement \citep{Nikolic2004}.
There is no dependence of the sSFR enhancement on the orbital directions
\citep{Keel1993}.

For minor mergers ($\mu > 3$), 
\citet{Woods2007} (see also \citealt{Li2008} and \citealt{Ellison2008})
found that only low mass late-type galaxies
(most being secondaries) in close pairs have significantly
enhanced sSFR while massive galaxies (most being primaries) show no significant
sSFR enhancement. This is exactly opposite to what \citet{Xu2010} found
for major mergers. The difference is likely due to the asymmetry of the 
gravitational effects in a minor merger (in particular when
$\mu$ is very high): 
for the massive primary the effects are minimal even when the two galaxies are
close while for the low mass secondary the effects can be overwhelming.

\vspace{2mm}
\noindent{\bf Summary on merger-induced sSFR enhancement (or lack of it):}
\begin{description}
\vspace{-2mm}
\item{1)} Statistically, only close mergers with separation 
$\rm r \leq 20\; h^{-1}kpc$ show significantly enhanced sSFR,
while wider mergers do not.
\vspace{-2mm}
\item{2)} Paired ellipticals are usually SFR quiet, same
as single ellipticals.
\vspace{-2mm}
\item{3)} {\bf For close major mergers, only massive SFGs} ($\rm M_{star}
  \gsim 10^{10.0}\; M_\solar$ for a Chabrier IMF) {\bf in SFG+SFG 
pairs}\footnote{SFGs do not include the ``red-and-dead'' 
S0/Sa galaxies among morphologically 
classified spirals.}{\bf in the field have significant sSFR enhancement 
(a factor of $\sim 3$). 
They are $\rm \sim 25\%$ of all spirals in close major mergers.}
\vspace{-2mm}
\item{4)} Spirals in close major mergers
are not sSFR enhanced when they are in one of the following categories:
low mass spirals, spirals in S+E pairs (``E'' including red S0/Sa),
spirals in dense-groups/clusters, spirals in pairs with 
$\rm \Delta v\gsim 400\; km\; sec^{-1}$.
\vspace{-6 mm}
\item{5)} For close minor mergers, only the low mass secondary spirals
are sSFR enhanced while the high mass primaries are not.
\vspace{-2 mm}
\end{description}

\bibliography{/Volumes/Seagate/data1/bibliography/ckxu_biblio}

\end{document}